\documentclass[12pt,preprint]{aastex}
\usepackage{amssymb}
\usepackage{graphicx}

\begin{document}

\title{Measuring Dark Energy with Gamma-Ray Bursts and Other Cosmological Probes}

\author{F. Y. Wang and Z. G. Dai}
\affil{Department of Astronomy, Nanjing University, Nanjing 210093, China}

\and

\author{Zong-Hong Zhu}
\affil{Department of Astronomy, Beijing Normal University,
                Beijing 100875, China}

\altaffiltext{}{E-mails: wfytxh2002@gmail.com (FYW), dzg@nju.edu.cn
(ZGD), zhuzh@bnu.edu.cn (ZHZ)}

\begin{abstract}
It has been widely shown that the cosmological parameters and dark
energy can be constrained by using data from type-Ia supernovae (SNe
Ia), the cosmic microwave background (CMB) anisotropy, the baryon
acoustic oscillation (BAO) peak from Sloan Digital Sky Survey
(SDSS), the X-ray gas mass fraction in clusters, and the linear
growth rate of perturbations at $z=0.15$ as obtained from the 2dF
Galaxy Redshift Survey. Recently, gamma-ray bursts (GRBs) have also
been argued to be promising standard candles for cosmography. In
this paper, we present constraints on the cosmological parameters
and dark energy by combining a recent GRB sample including 69 events
with the other cosmological probes. First, we find that for the
$\Lambda$CDM cosmology this combination makes the constraints
stringent and the best fit is close to the flat universe. Second, we
fit the flat Cardassian expansion model and find that this model is
consistent with the $\Lambda$CDM cosmology. Third, we present
constraints on several two-parameter dark energy models and find
that these models are also consistent with the $\Lambda$CDM
cosmology. Finally, we reconstruct the dark energy equation-of-state
parameter $w(z)$ and the deceleration parameter $q(z)$. We see that
the acceleration could have started at a redshift from
$z_T=0.40_{-0.08}^{+0.14}$ to $z_T=0.65_{-0.05}^{+0.10}$. This
difference in the transition redshift is due to different dark
energy models that we adopt. The most stringent constraint on $w(z)$
lies in the redshift range $z\sim0.3-0.6$.
\end{abstract}


\section{Introduction}\label{sec:introduction}
The traditional cosmology has been revolutionized by modern
observational techniques in distant Type Ia supernovae (SNe Ia)
(Riess et al. 1998; Perlmutter et al. 1999), cosmic microwave
background (CMB) fluctuations (Bennett et al. 2003; Spergel et al.
2003, 2006), and large-scale structure (LSS) (Tegmark et al. 2006).
These observations suggest that the composition of the universe may
consist of an extra component such as dark energy or the equations
governing gravity may need a variation to explain the acceleration
of the universe at the present epoch.

SNe Ia have been considered as astronomical standard candles and
used to measure the geometry and dynamics of the universe. However,
since it is difficult to observe SNe Ia at redshift $z\gtrsim 1.7$,
this measurement has been carried out only for the $z\lesssim 1.7$
universe. Recently, it was shown that GRBs may be complementary to
the SN cosmology for three reasons. First, GRBs are the most
powerful explosive events at cosmological distances and in
particular long-duration GRBs originate from the core collapse of
massive stars. So GRBs would be detectable out to very high
redshifts when the core collapse of the first stars occur (Ciardi \&
Loeb 2000; Lamb \& Reichart 2000; Bromm \& Loeb 2002, 2006). In
fact, the farthest burst detected so far is GRB 050904, which is at
$z=6.295$ (Kawai et al. 2006). Thus, GRBs could provide a much
longer arm for measuring changes in the slope of the Hubble diagram
than do SNe Ia. Second, gamma-ray photons suffer from no dust
extinction when they propagate to us, so the observed gamma-ray flux
is a direct measurement of the prompt emission energy. Third, there
have been extensive discussions on relations between the spectral
and temporal properties and some of these relations have been shown
to be promising standard candles for cosmography. Schaefer (2003)
derived the luminosity distances of 9 GRBs with known redshifts by
using two quantities (the spectral lag and the variability) as
luminosity calibrators and gave a constraint on the mass density
$\Omega_M$. Ghirlanda et al. (2004a) found a tight relation between
collimation-corrected energy $E_{\gamma}$ and the local-observer
peak energy $E_{p}^{'}$ (i.e., the so-called Ghirlanda relation).
This relation may be physically understood as due to the viewing
angle effect of an annular jet (Levinson \& Eichler 2005) or
Comptonization of the thermal radiation flux that is advected from
the base of an outflow (Rees \& \& M\'esz\'aros 2005; Thompson et
al. 2006). Assuming that some physical explanation (e.g., the
understandings mentioned above) comes into existence, Dai, Liang \&
Xu (2004) used the Ghirlanda relation to constrain the cosmological
parameters and dark energy. Since then, a lot of work in this
so-called {\em GRB cosmology} field has been published (Ghirlanda et
al. 2004b; Di Girolamo et al. 2005; Firmani et al. 2005; Friedman \&
Bloom 2005; Lamb et al. 2005; Liang \& Zhang 2005, 2006; Xu, Dai \&
Liang 2005; Wang \& Dai 2006a; Li et al. 2006; Su et al. 2006;
Schaefer 2007; Wright 2007). Very recently, Schaefer (2007) used 69
GRBs and five relations to build the Hubble diagram out to $z=6.60$
and discussed the properties of dark energy in several dark energy
models. He found that the GRB Hubble diagram is consistent with the
concordance cosmology. Besides SNe Ia and GRBs, the other
observations such as the shift parameter of CMB (Spergel et al.
2003, 2006), the baryon acoustic peak from Sloan Digital Sky Survey
(SDSS) (Eisenstein et al. 2005), the X-ray gas mass fraction in
clusters (Allen et al. 2004), the perturbation growth rate from 2dF
Galaxy Redshift Survey (Hawkins et al. 2003), and the weak lensing
(e.g., Schimd et al. 2007) have been used to constrain cosmological
parameters and explore the properties of dark energy.

It is of growing interest that dark energy is reconstructed in a
model-independent way to investigate the evolution of the
deceleration parameter $q(z)$ and the dark-energy equation-of-state
parameter $w(z)$ (Alam et al. 2004; Virey et al. 2005; Gong \& Wang
2007; Alam et al; 2007). Evolving dark energy models had been shown
to satisfy the data from SNe Ia. To reconstruct $q(z)$ and $w(z)$,
Gong \& Wang (2007) used the new ``Gold" sample of SNe and data of
SDSS and CMB, while Alam et al. (2007) adopted the new ``Gold" SN
sample, the SNLS sample, and data of SDSS and CMB. It is found that
the result is strongly dependent on the matter density $\Omega_{M}$.
The transition redshift $z_T\sim0.2$ was found in reconstruction of
$q(z)$ (Virey et al. 2005; Shapiro \& Turner 2006; Gong \& Wang
2006). Previous investigations in the construction of $w(z)$ show
that the stringent constraint on $w(z)$ is in the redshift range
$z\sim0.2-0.5$ (Alam et al. 2004; Gong \& Zhang 2005).

In this paper we use GRBs and the other observational data to
measure the cosmological parameters and the nature of dark energy.
We also reconstruct $q(z)$ and $w(z)$ out to $z>6.0$ using these
observational datasets, explore the transition redshift and
constrain $w(z)$. Recently, Su et al. (2006), Li et al. (2006), and
Wright (2007) combined GRBs with some other cosmological probes to
constrain the $\Lambda$CDM cosmology, the constant $w$ model, and
the dark energy model of $w(z)=w_0+w_a(1-a)$ (where $a$ is the scale
factor of the universe), respectively. In their papers, these
authors adopted the distance modulus and its error (of a GRB)
calculated for the concordance cosmology or the dynamical dark
energy model of $w(z)=-1.31+1.48z$, which were presented by Schaefer
(2007). In addition, Li et al. (2006) used the Markov Chain Monte
Carlo technique to carry out global fitting. Here we use the
observational data (e.g., time lag, variability, spectral peak
energy $E_{\rm peak}$, minimum rise time) of GRBs to make a
simultaneous fit of five correlations in any given cosmology, and
consider more other cosmological probes and more dark energy models.
The structure of this paper is arranged as follows: in section 2, we
introduce GRBs and the other cosmological probes and describe our
analytical methods. The constraints on the cosmological parameters
and dark energy are presented in section 3. In section 4, we
reconstruct $w(z)$ and $q(z)$. In sections 5, we summarize our
findings and present a brief discussion.

\section{Observational data and Analysis Methods}\label{sec:analysis}
\subsection{Type Ia Supernovae (SNe Ia)}
Riess et al. (2004) reanalyzed the SN Ia dataset. They considered 14
new high-redshift events observed by the Hubble Space Telescope
(HST). This led to a sample known as the ``Gold" sample containing
157 SNe Ia. Recently, Riess et al. (2007) added $25$ SNe Ia to this
sample. The final sample now consists of $182$ SNe Ia. The
observations of SNe Ia provide the currently most direct way of
probing the dark energy at low-to-medium redshifts because the used
luminosity distance is directly related to the expansion history of
the universe, that is,
\begin{equation}
d_{L}=\left\{
\begin{array}{l}
\displaystyle
cH_{0}^{-1}(1+z)(-\Omega_{k})^{-1/2}\sin[(-\Omega_{k})^{1/2}I]
\phantom{sssssssssssssssss}  \Omega_{k}<0, \\
\displaystyle cH_{0}^{-1}(1+z)I
\phantom{sssssssssssssssssssssssssssssssssssssss}
\Omega_{k}=0,\\
\displaystyle
cH_{0}^{-1}(1+z)\Omega_{k}^{-1/2}\sinh[\Omega_{k}^{1/2}I]
\phantom{ssssssssssssssssssssssss}\,  \Omega_{k}>0,\\
\end{array} \right.
\label{eqn:fc:}
\end{equation}
where
\begin{equation}
\Omega_{k}=1-\Omega_{M}-\Omega_{DE},
\end{equation}
\begin{equation}
I=\int_{0}^{z}dz/E(z),
\end{equation}
\begin{equation}
E(z)=[(1+z)^{3}\Omega_{M}+f(z)\Omega_{DE}+(1+z)^{2}\Omega_{k}]^{1/2},
\end{equation}
\begin{equation}
f(z)=\exp\left[3\int_{0}^{z}\frac{(1+w(z'))dz'}{(1+z')}\right],
\end{equation}
where $w(z)$ is the equation-of-state parameter for dark energy and
$d_{L}$ is the luminosity distance. With $d_{L}$ in units of
megaparsecs, the predicted distance modulus is
\begin{equation}
\mu=5\log(d_{L})+25.
\end{equation}
The likelihood functions for the parameters $\Omega_{M}$ and
$\Omega_{DE}$ can be determined from $\chi^{2}$ statistics,
\begin{equation}
\chi^{2}(H_{0},\Omega_{M},\Omega_{DE})=\sum_{i=1}^{N}
\frac{[\mu_{i}(z_{i},H_{0},\Omega_{M},\Omega_{DE})-\mu_{0,i}]^{2}}
{\sigma_{\mu_{0,i}}^{2}+\sigma_{\nu}^{2}},
\end{equation}
where $\sigma_{\nu}$ is the dispersion in the supernova redshift
(transformed to distance modulus) due to a peculiar velocity,
$\mu_{0,i}$ is the observed distance modulus, and
$\sigma_{\mu_{0,i}}$ is the uncertainty in the individual distance
modulus. The confidence regions in the $\Omega_{M}-\Omega_{DE}$
plane can be found through marginalizing the likelihood functions
over $H_{0}$ (i.e., integrating the probability density
$p\propto\exp(-\chi^2/2)$ for all values of $H_{0}$).

\subsection{Gamma-Ray Bursts (GRBs)}
GRBs can be detected out to very high redshifts (Ciardi \& Loeb
2000; Lamb \& Reichart 2000; Bromm \& Loeb 2002, 2006). They can
bridge up the gap between the nearby SNe Ia and the distant CMB
anisotropy. Schaefer (2007) complied 69 GRBs to make simultaneous
uses of five luminosity indicators, which are relations of
$\tau_{\rm lag}-L$, $V-L$, $E_{\rm peak}-L$, $E_{\rm
peak}-E_{\gamma}$, and $\tau_{\rm RT}-L$. Here the time lag
($\tau_{\rm lag}$) is the time shift between the hard and soft light
curves, $L$ is the luminosity of a GRB, the variability $V$ of a
burst denotes whether its light curve is spiky or smooth and $V$ can
be obtained by calculating the normalized variance of an observed
light curve around a smoothed version of that light curve (Fenimore
\& Ramirez- Ruiz 2000), $E_{\rm peak}$ is the peak energy in the
$\nu F_{\nu}$ spectrum, $E_{\gamma}=(1-\cos\theta_j)E_{\rm iso}$ is
the collimation-corrected energy of a GRB, and the minimum rise time
($\tau_{\rm RT}$) in the gamma-ray light curve is the shortest time
over which the light curve rises by half of the peak flux of the
pulse. We make a simultaneous fit to these five relations for any
fixed cosmology. We perform a linear regression analysis to find a
relation between observational quantities. After obtaining the
distance modulus of each burst using one of these relations, we use
the same method as Schaefer (2007) to calculate the real distance
modulus,
\begin{equation}
\mu_{\rm fit}=(\sum_i \mu_i/\sigma_{\mu_i}^2)/(\sum_i
\sigma_{\mu_i}^{-2}),
\end{equation}
where the summation runs from $1-5$ over the relations with
available data, $\mu_i$ is the best estimated distance modulus from
the $i$-th relation, and $\sigma_{\mu_i}$ is the corresponding
uncertainty. The uncertainty of the distance modulus for each burst
is
\begin{equation}
\sigma_{\mu_{\rm fit}}=(\sum_i \sigma_{\mu_i}^{-2})^{-1/2}.
\end{equation}
Fig.1 shows the Hubble diagram from the new ``Gold" SNIa sample and
69 GRBs. The combined Hubble diagram is consistent with the
concordance cosmology. GRBs can build the Hubble diagram out to
$z>6.0$ (Schaefer 2007). The GRB Hubble diagram is well-behaved and
describes the shape of the Hubble diagram at high redshifts. When
calculating constraints on cosmological parameters and dark energy,
we do not care about the slopes of the five relations because we
have marginalized these parameters (Schaefer 2007). The
marginalization method is to integrate over some parameter for all
of its possible values. We also marginalize the nuisance parameter
$H_{0}$. The $\chi^2$ value is
\begin{equation}
\chi^{2}(H_{0},\Omega_{M},\Omega_{DE})=\sum_{i=1}^{N}
\frac{[\mu_{i}(z_{i},H_{0},\Omega_{M},\Omega_{DE})-\mu_{{\rm
fit},i}]^{2}}{\sigma_{\mu_{{\rm fit},i}}^{2}},
\end{equation}
where $\mu_{{\rm fit},i}$ and $\sigma_{\mu_{{\rm fit},i}}$ are the
fitted distance modulus and its error.

\subsection{Cosmic Microwave Background (CMB)}
Observations of the CMB anisotropy provide us with very accurate
measurements, which may be used to gain insight about dark energy
and cosmological parameters (Spergel et al. 2006). We may make use
of the 3-year WMAP results to get the shift parameter (Wang \&
Mukherjee 2006)
\begin{equation}
\label{shift}
\mathcal{R}=\frac{\sqrt{\Omega_M}}{\sqrt{|\Omega_k|}}{\rm
sinn}\left(\sqrt{|\Omega_k|}\int_{0}^{z_{\rm
ls}}\frac{dz}{E(z)}\right)=1.70\pm 0.03,
\end{equation}
where $E(z) \equiv H(z)/H_0$ and the function ${\rm sinn}(x)$ is
defined as ${\rm sinn}(x) = \sin(x)$ for a closed universe, ${\rm
sinn}(x) = \sinh(x)$ for an open universe and ${\rm sinn}(x) = x$
for a flat universe. To calculate the last scattering redshift
$z_{\rm ls}$, we adopt $\Omega_b h^2=0.024$ and $\Omega_{M}h^2 =0.14
\pm 0.02$. To calculate $z_{\rm ls}$, we consider a fitting
function:
\begin{equation}
z_{\rm ls}=1048[1+0.00124 (\Omega_{b}h^2)^{-0.738}][1+g_1
(\Omega_{M}h^2)^{g_2}],
\end{equation}
where the quantities $g_1$ and $g_2$ are defined as $
g_1=0.078(\Omega_b h^2)^{-0.238} [1+39.5 (\Omega_b
h^2)^{0.763}]^{-1}$ and $g_2=0.56 [1+21.1 (\Omega_b
h^2)^{1.81}]^{-1}$ respectively (Hu \& Sugiyama 1996).  The $\chi^2$
value is
\begin{equation}
\chi^{2}_{\rm CMB}=\frac{(\mathcal{R}-1.70)^2}{0.03^2}.
\end{equation}

\subsection{Baryon Acoustic Peak from SDSS}
It is well known that the acoustic peaks in the CMB anisotropy power
spectrum can be used to determine the properties of perturbations
and to constrain cosmological parameters and dark energy (Spergel et
al. 2003). The acoustic peaks occur because the cosmic perturbations
excite sound waves in the relativistic plasma of the early universe
(Peebles \& Yu 1970; Holtzmann 1989). Because the universe has a
fraction of baryons, the acoustic oscillations in the relativistic
plasma would be imprinted onto the late-time power spectrum of the
nonrelativistic matter (Peebles \& Yu 1970; Eisenstein \& Hu 1998).
The acoustic signatures in the large-scale clustering of galaxies
can also be used to constrain cosmological parameters and dark
energy by detection of a peak in the correlation function of
luminous red galaxies in the SDSS (Eisenstein et al. 2005). This
peak can provide a ``standard ruler" with which the cosmological
parameters and dark energy are measured. We use the value
\begin{equation}
A = \frac{\sqrt{\Omega_{M}}}{z_1}
 \left[\frac{z_1}{E(z_1)}\frac{1}{|\Omega_k|} {\rm sinn}^2
 \left(\sqrt{|\Omega_k|}\int_0^{z_1}\frac{dz}{E(z)}\right)\right]^{1/3},
\end{equation}
measured from the SDSS data to be $A=0.469(0.95/0.98)^{-0.35}\pm
0.017$, where $z_1 = 0.35$. The $\chi^2$ value is
\begin{equation}
\chi^{2}_{\rm BAO}=\frac{(A-0.469)^2}{0.017^2}.
\end{equation}

\subsection{X-ray Gas Mass Fraction in Clusters}
Since clusters of galaxies are the largest virialized systems in the
universe, their matter content is thought to provide a sample of the
matter content of the universe. A comparison of the gas mass
fraction, $f_{\rm gas} = M_{\rm gas} / M_{\rm tot}$, as inferred
from X-ray observations of clusters of galaxies to the cosmic baryon
fraction can provide a direct constraint on the density parameter of
the universe $\Omega_M$ (White et al. 1993). Moreover, assuming the
gas mass fraction is constant in cosmic time, Sasaki (1996) showed
that the $f_{\rm gas}$ measurements of clusters of galaxies  at
different redshifts also provide an efficient way to constrain other
cosmological parameters decribing the geometry of the universe. This
is based on the fact that the measured $f_{\rm gas}$ values for each
cluster of galaxies depend on the assumed angular diameter distances
to the sources as $f_{\rm gas} \propto (D^A)^{3/2}$. The true,
underlying cosmology should be the one which makes these measured
$f_{\rm gas}$ values invariant with redshift (Sasaki 1996; Allen at
al. 2004). Using the {\it Chandra} observational data, Allen et al.
(2004) have got the $f_{\rm gas}$ profiles for the 26 relaxed
clusters. These authors used the 26-cluster data to constrain
cosmological parameters. They found $\Omega_{M} =
0.245^{+0.040}_{-0.037}$ and $\Omega_{\Lambda} =
0.96^{+0.19}_{-0.22}$ in the $\Lambda$CDM cosmology. This database
has also been used to constrain the generalized Chaplygin gas model
(Zhu 2004) and the braneworld cosmology (Zhu and Alcaniz 2005). We
will combine this probe in our analysis. Following Allen et al.
(2004), we calculate the $\chi^2$ value as
\begin{eqnarray}
\chi^2_{\rm gas}& =& \left( \sum_{i=1}^{26} \frac{\left[f_{\rm
gas}^{\rm SCDM}(z_{\rm i})- f_{\rm gas,\,i}
\right]^2}{\sigma_{f_{\rm gas,\,i}}^2}\nonumber \right)
\end{eqnarray}

\begin{eqnarray}
+\left(\frac{\Omega_{\rm b}h^2-0.0233}{0.0008} \right)^2
+\left(\frac{h-0.72} {0.08} \right)^2+\left(\frac{b-0.824} {0.089}
\right)^2,
\end{eqnarray}
where $f_{\rm gas}^{\rm
SCDM}(z)=b\Omega_b/[(1+0.19\sqrt{h})\Omega_M]\times [d_A^{\rm
SCDM}(z)/d_A^{\rm mod}(z)]^{1.5}$, $f_{{\rm gas},i}$ is the
observational baryon gas mass fraction and $b$ is a bias factor
motivated by gas-dynamical simulations which suggest that the baryon
fraction in clusters is slightly lower than for the universe as a
whole.

\subsection{Perturbation Growth Rate from 2dF Galaxy Redshift Survey}
The clustering of galaxies is determined by the initial mass
fluctuations and their evolution. We can set constraints on the
initial mass fluctuations and their evolution by measuring the
galactic two-point correlation function. The 2dF galaxy redshift
survey measured the two point correlation function at the redsift of
$z=0.15$. Hawkins et al. (2003) measured the redshift distortion
parameter $\beta=0.49\pm0.09$. This result can be combined with the
linear bias parameter ${\bar b}=1.04\pm0.11$. So the growth factor
$g$ at $z= 0.15$ is $g={\bar b}\times \beta=0.51\pm0.11$.
Theoretically, this growth factor is cosmology-dependent. Thus, the
measurement of the perturbation growth rate (PGR) $g(z=0.15)$ can be
used to calculate $\chi^{2}$:
\begin{equation}
\chi^{2}_{\rm PGR}=\frac{(g-0.51)^2}{0.11^2},
\end{equation}
which constrains the cosmological parameters and dark energy.

\section{Constraints on Cosmological Parameters and Dark Energy}
Using the datasets of the above observational techniques, we measure
cosmological parameters and dark energy. We can combine these probes
by multiplying the likelihood functions. The total $\chi^2$ value is
\begin{equation}
\chi^2_{\rm total}=\chi^2_{\rm SN}+\chi^2_{\rm GRB}+\chi^2_{\rm
CMB}+\chi^2_{\rm BAO}+\chi^2_{\rm gas}+\chi^2_{\rm PGR}
\end{equation}

\subsection{The $\Lambda$CDM Cosmology}
The luminosity distance in a Friedmann-Robertson-Walker (FRW)
cosmology with mass density $\Omega_M$ and vacuum energy density
(i.e., the cosmological constant) $\Omega_\Lambda$ is (Carroll,
Press \& Turner 1992)
\begin{eqnarray}
d_L & = & c(1+z)H_0^{-1}|\Omega_k|^{-1/2}{\rm
sinn}\{|\Omega_k|^{1/2}\nonumber
\\ & & \times
\int_0^zdz[(1+z)^2(1+\Omega_Mz)-z(2+z)\Omega_\Lambda]^{-1/2}\}.
\end{eqnarray}
We use the datasets discussed above to constrain cosmological
parameters. Fig.2 shows the $1\sigma$ contours plotting in the
$\Omega_{M}-\Omega_{\Lambda}$ plane. The thick black line contour
from all the datasets shows $\Omega_{M}=0.27\pm0.02$ and
$\Omega_{\Lambda}=0.73\pm0.08$ ($1\sigma$) with $\chi^{2}_{\rm
min}=270.60$. The red contour shows a constraint from 69 GRBs, and
for a flat universe, we measure $\Omega_M=0.34_{-0.10}^{+0.09}$
($1\sigma$), which is consistent with Schaefer (2007). Because the
thin solid line in Fig.2 represents a flat universe, our result from
all the datasets favors a flat universe.

\subsection{The Cardassian Expansion Model}
The Cardassian expansion models (Freese \& Lewis 2002) involve a
modification of the Friedmann equation, which allows an acceleration
in a flat, matter-dominated cosmology. We assume that the Cardassian
expansion model is (Freese \& Lewis 2002; Zhu et al. 2004)
\begin{equation}
H^2=\frac{8\pi G}{3}(\rho+C\rho^n).
\end{equation}
This modification may arise from embedding our observable universe
as a (3+1)-dimensional brane in extra dimensions or the
self-interaction of dark matter. The luminosity distance in this
model is
\begin{equation}
d_{L}=cH_{0}^{-1}(1+z)\int_{0}^{z}dz[(1+z)^{3}\Omega_{M}
+(1-\Omega_{M})(1+z)^{3n}]^{-1/2}.
\end{equation}
Fig.3 shows constraints on $\Omega_{M}$ and $n$. The solid contours
are obtained from all the datasets. From this figure, we have
$\Omega_{M}=0.28\pm0.02$ and $n=0.02^{+0.10}_{-0.09}$ at the
$1\sigma$ confidence level with $\chi^2_{\rm min}=272.52$. This
result is consistent with the $\Lambda$CDM cosmology.

\subsection{The $w(z)=w_{0}$ Model}
We consider an equation of state for dark energy
\begin{equation}
w(z)=w_{0}.
\end{equation}
In this dark energy model, the luminosity distance for a flat
universe is (Riess et al. 2004)
\begin{equation}
d_{L}=cH_{0}^{-1}(1+z)\int_{0}^{z}dz[(1+z)^{3}\Omega_{M}+(1-\Omega_{M})
(1+z)^{3(1+w_{0})}]^{-1/2}
\end{equation}
Fig.4 shows the constraints on $w_{0}$ versus $\Omega_{M}$ in this
dark energy model from all the datasets. From this figure, we have
$\Omega_{M}=0.31\pm0.03$ and $w_{0}=-0.95^{+0.16}_{-0.13}$
($1\sigma)$ with $\chi^2_{\rm min}=272.23$.

\subsection{Two-Parameter Dark Energy Models}
Using the parameterization
\begin{equation}
w(z)=w_{0}+\frac{w_{1}z}{1+z},
\end{equation}
the luminosity distance is calculated by (Chevallier \& Polarski
2001; Linder 2003)
\begin{equation}
d_{L}=cH_{0}^{-1}(1+z)\int_{0}^{z}dz[(1+z)^{3}\Omega_{M}
+(1-\Omega_{M})(1+z)^{3(1+w_{0}+w_{1})}e^{-3w_{1}z/(1+z)}]^{-1/2}.
\end{equation}
Fig.5 shows the constraints on $w_{0}$ versus $w_{1}$ in this dark
energy model. The solid contours are obtained from all the datasets
and we find $\chi^{2}_{\rm min}=273.25$,
$w_{0}=-1.08_{-0.32}^{+0.20}$ and $w_{1}=0.84_{-0.82}^{+0.40}$
($1\sigma$) for the prior of $\Omega_{M}=0.30$. We also assume this
prior in the following analysis.

Jassal, Bagla and Padmanabhan (2004) modified the above
parameterization as
\begin{equation}
w(z)=w_{0}+\frac{w_{1}z}{(1+z)^2}.
\end{equation}
This equation can model a dark energy component which has a similar
value at lower and higher redshifts. The luminosity distance is
\begin{equation}
d_{L}=cH_{0}^{-1}(1+z)\int_{0}^{z}dz[(1+z)^{3}\Omega_{M}
+(1-\Omega_{M})(1+z)^{3(1+w_{0})}e^{3w_{1}z^{2}/2(1+z)^{2}}]^{-1/2}.
\end{equation}
Constraints on $w_{0}$ and $w_{1}$ are presented in Fig.6. From this
figure, we find $\chi^{2}_{\rm min}=272.07$,
$w_{0}=-1.36_{-0.48}^{+0.38}$ and $w_{1}=3.32_{-2.82}^{+2.28}$
($1\sigma$) from all the datasets (blue contour).

The third dark energy model that we consider is (Alam et al. 2003)
\begin{equation}
w(z)=\frac{1+z}{3}\frac{A_1+2A_2(1+z)}{\Omega_{DE}(z)}-1,
\end{equation}
where $\Omega_{DE}(z)$ is defined as
\begin{equation}
\Omega_{DE}(z)=A_1(1+z)+A_2(1+z)^2+1-\Omega_M-A_1-A_2.
\end{equation}
Fig.7 shows the constraints on $A_{1}$ versus $A_{2}$ in this dark
energy model. The solid contours are obtained from all the datasets
and we find $\chi^{2}_{\rm min}=273.95$,
$A_{1}=-0.43_{-1.08}^{+0.96}$ and $A_{2}=0.22_{-0.32}^{+0.29}$
($1\sigma$).

\section{Reconstruction of $w(z)$ and $q(z)$}\label{recs}
Many dark energy models have been proposed (Copeland et al. 2006;
Bludman 2006 for a recent review) and we have fitted these models
using the observational data in the last section. We now explore the
properties of dark energy in a model-independent way (Sahni et al.
2006 for a review). In the following we reconstruct dark energy to
find new information about dark energy from most of the recent
datasets. The method to reconstruct directly properties of dark
energy from observations in a quasi-model independent method has
been discussed (Alam et al. 2004; Gong \& Wang. 2007; Alam et al.
2007). We determine the dark energy equation of state based on
\begin{equation}
w(z)=\frac{\frac{2}{3}(1+z)\frac{d\ln
H}{dz}-1}{1-\Omega_{M}H^{-2}(1+z)^3}.
\end{equation}
The deceleration parameter
\begin{equation}
q(z)=(1+z)H^{-1}\frac{dH}{dz}-1.
\end{equation}
We consider the first ansatz
\begin{equation}
H(z)=H_{0}[(1+z)^{3}\Omega_{M}+(1-\Omega_{M})(1+z)^{3(1+w_{0}
+w_{1})}e^{-3w_{1}z/(1+z)}]^{1/2}],
\end{equation}
which is in fact equivalent to the parameterization equation (24).
The evolution of $w(z)$ is plotted in Fig.8. It is easy to see that
the errors of the constraint on the equation of state become larger
with redshift. The stringent constraint on $w(z)$ happens at
$z=0.3\sim0.7$. Using the GRB data, we can reconstruct $w(z)$ out to
$z\sim6.0$ in the bottom panel. The evolution of $q(z)$ is plotted
in Fig.9. We can see that the transition redshift at which the
expansion of the universe was from deceleration $(q(z)>0)$ to
acceleration $(q(z)<0)$ is $z_{T}=0.57_{-0.07}^{+0.08}$ ($1\sigma$).
This result is consistent with Riess et al. (2004) and Wang \& Dai
(2006a, 2006b).

We consider the second ansatz
\begin{equation}
H(z)=H_{0}[(1+z)^{3}\Omega_{M}+(1-\Omega_{M})(1+z)^{3(1+w_{0})}
e^{3w_{1}z^{2}/2(1+z)^{2}}]^{1/2},
\end{equation}
which is in fact equivalent to the parameterization equation (26).
The evolution of $w(z)$ is plotted in Fig.10. The stringent
constraint on $w(z)$ happens at $z=0.2\sim0.35$. Using the GRB data,
we can reconstruct $w(z)$ out to $z\sim6.0$ in the bottom panel. We
find that the constraint on $w(z)$ is also stringent around
$z=4.0\sim5.0$. The evolution of $q(z)$ is plotted in Fig.11. We can
see that the transition redshift is $z_T=0.40_{-0.08}^{+0.14}$
($1\sigma$).

We consider the third ansatz
\begin{equation}
H(z)=H_{0}[(1+z)^{3}\Omega_{M}+A_1(1+z)+A_2(1+z)^2+1-\Omega_M-A_1-A_2]^{1/2},
\end{equation}
which is in fact equivalent to the parameterization equation (28).
The evolution of $w(z)$ is plotted in Fig.12. The stringent
constraint on $w(z)$ happens at $z=0.35\sim0.55$. Using the GRB
data, we can reconstruct $w(z)$ out to $z\sim6.0$ in the bottom
panel. We find that the constraint on $w(z)$ becomes stringent
around $z\sim6.0$. The evolution of $q(z)$ is plotted in Fig.13.
>From this figure, we can see that the transition redshift is
$z_T=0.65_{-0.05}^{+0.10}$ ($1\sigma$).

\section{Conclusions and Discussion}\label{discussion}
In this paper, we have presented the constraints on the cosmological
parameters and dark energy by combining a recent GRB sample
including 69 events with the 182 SNe Ia, CMB, BAO, the X-ray gas
mass fraction in clusters and the linear growth rate of
perturbations at $z=0.15$ as obtained from the 2dF galaxy redshift
survey. We found that the mass density of the universe is
$\Omega_{M}=0.27\pm0.02$ and $\Omega_{\Lambda}=0.73\pm0.08$
($1\sigma$) in the $\Lambda$CDM cosmology. This result is well
consistent with a flat universe. We also found that
$\Omega_{M}=0.28\pm0.02$ and $n=0.02^{+0.10}_{-0.09}$ ($1\sigma$) in
the flat Cardassian expansion model. We fitted several dark energy
models. Finally, we reconstructed the dark energy equation-of-state
parameter $w(z)$ and the deceleration parameter $q(z)$. We found
that the the cosmic acceleration could have started between the
redshift $z_T=0.40_{-0.08}^{+0.14}$ and $z_T=0.65_{-0.05}^{+0.10}$
($1\sigma$). The stringent constraints on $w(z)$ lie in the redshift
range $z\sim0.3-0.6$.

Based on our analysis, it can be seen that the preferred
cosmological model is the flat $\Lambda$CDM cosmology because of a
small minimum $\chi^2$ value, $\chi^2_{\rm min}=270.60$. The other
models such as the Cardassian expansion model, the flat constant $w$
model, and three two-parameter dark energy models can also fit all
the datasets because the minimum $\chi^2$ values in these models
vary only from $\chi^2_{\rm min}=272.23$ to $\chi^2_{\rm
min}=273.95$. Thus, we cannot reject any of these models.

It is well known that the cosmological constant suffers from the
``fine tuning" problem and the coincidence problem (Zeldovich 1968;
Weinberg 1989). In this paper, therefore, we have considered
alternative possibilities, e.g., the Cardassian expansion model, the
flat constant $w$ model, and three two-parameter dark energy models.
As we have shown, all the alternative models can be reduced to the
flat $\Lambda$CDM cosmology at the $1\sigma$ confidence level. So
one needs more new observed data to distinguish between these
models. New observations would be expected to improve the current
constraints and test the flat $\Lambda$CDM model. GRBs appear to be
natural events to study the universe at very high redshifts. The
forthcoming GLAST will accumulate more GRB data, and in particular,
its combination with {\em Swift} would lead to stronger constraints
on high-redshift properties of dark energy.

\acknowledgments We thank the referee for his/her detailed and very
constructive suggestions that have allowed us to improve our
manuscript. This work is supported by the National Natural Science
Foundation of China (grants 10221001 and 10640420144) and the
Scientific Research Foundation of Graduate School of Nanjing
University(for FYW). ZHZ acknowledges support from the National
Natural Science Foundation of China, under Grant No. 10533010, and
SRF for ROCS, SEM of China.

\begin{figure}
\begin{center}
\includegraphics[angle=0,width=0.5\textwidth]{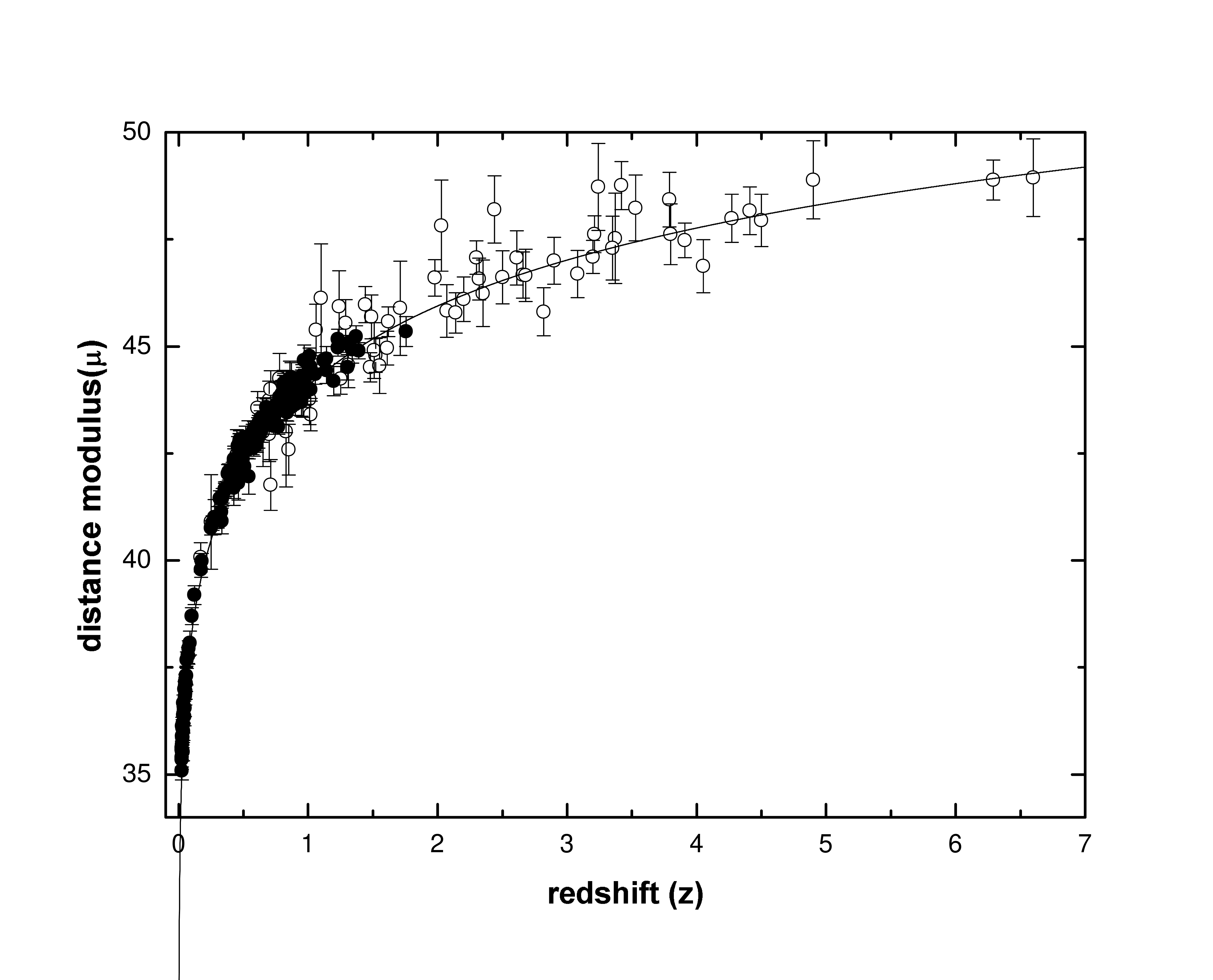}
\caption{Hubble diagram of new 182 SNe Ia (filled circles) and 69
GRBs (open circles). The solid line is calculated for a flat
cosmology: $\Omega_{M}=0.27$ and $\Omega_{\Lambda}=0.73$.
 \label{fig1}}
\end{center}
\end{figure}

\clearpage

\begin{figure}
\includegraphics[angle=0,width=0.5\textwidth]{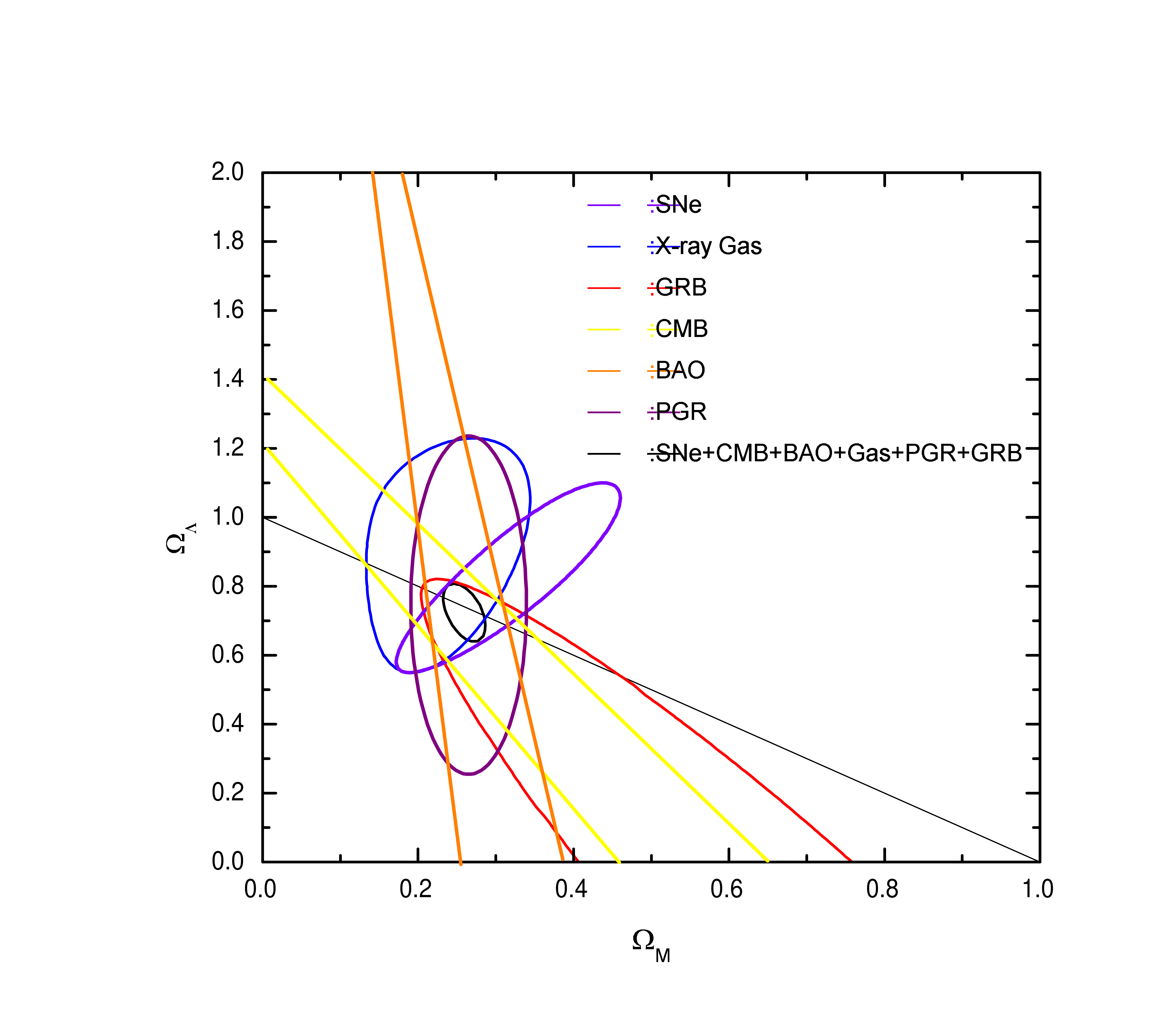} \caption{The $1\sigma$ joint confidence contours
for $(\Omega _M,\Omega _\Lambda)$ from the observational datasets.
The thick black line contour corresponds to all the datasets. The
blue contour corresponds to 26 galaxy clusters. The red contour
corresponds to 69 GRBs. The yellow contour corresponds to the CMB
shift parameter. The violet contour corresponds to 182 SNe Ia. The
orange contour corresponds to BAO. The purple contour corresponds to
2dF Galaxy Redshift Survey. The thin solid line represents a flat
universe \label{fig2}}
\end{figure}

\begin{figure}
\begin{center}
\includegraphics[angle=0,height=1.0\textheight,width=1.0\textwidth]{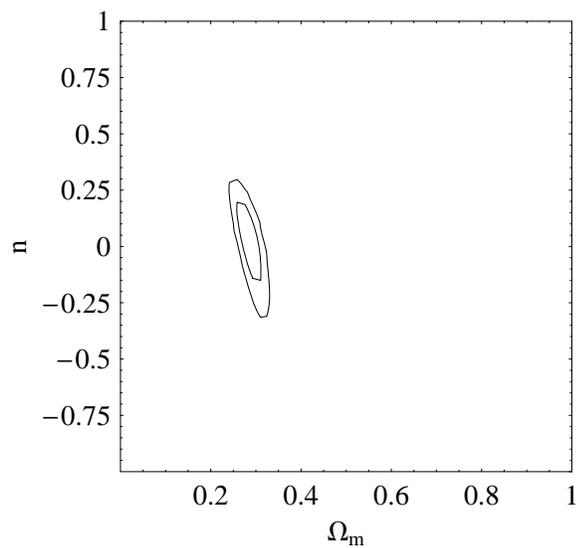}
\caption{The $1\sigma$ and $2\sigma$ joint confidence contours for
$(\Omega_{M}, n)$ from all the observational data.
 \label{fig3}}
\end{center}
\end{figure}

\begin{figure}
\plotone{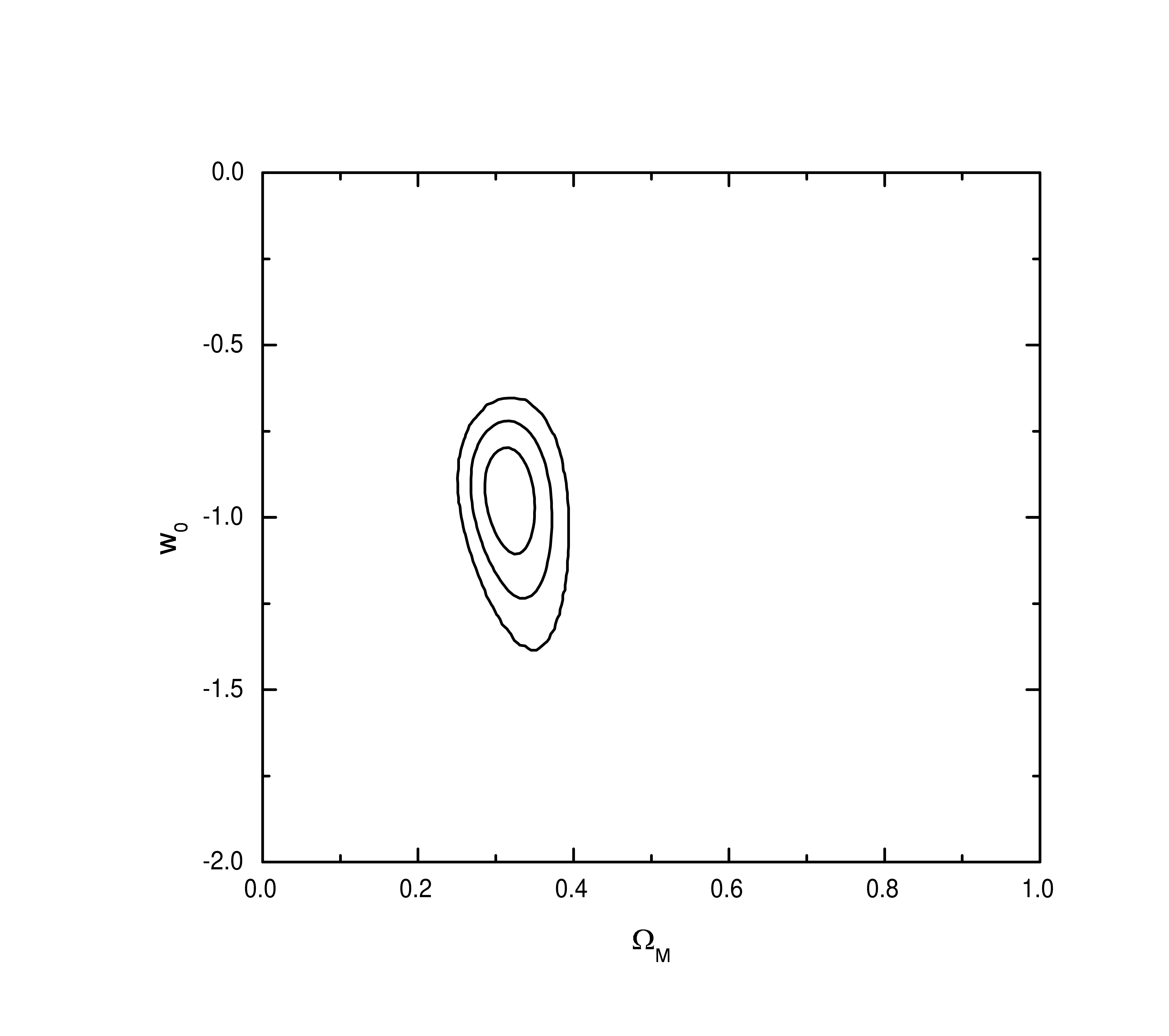} \caption{The $1\sigma$, $2\sigma$ and $3\sigma$
joint confidence contours for $(\Omega _M ,w_{0})$ from all the
observational datasets. \label{fig4}}
\end{figure}

\begin{figure}
\begin{center}
\includegraphics[angle=0,height=1.0\textheight,width=1.0\textwidth]{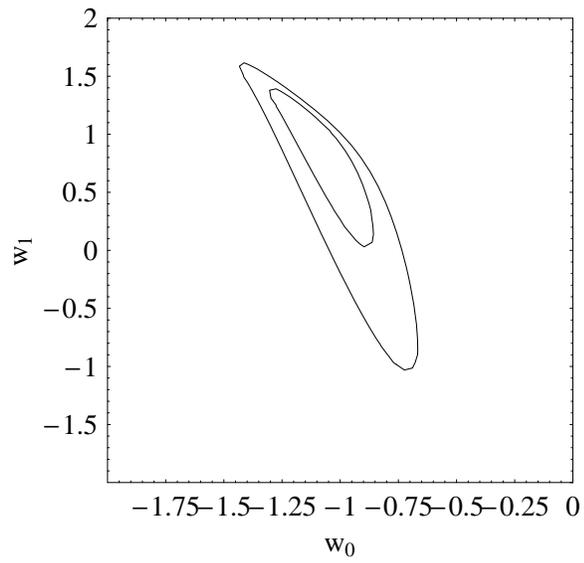}
\caption{The $1\sigma$ and $2\sigma$ joint confidence contours of
from all the observational data in the $w(z)=w_{0}+w_{1}z/(1+z)$
model. \label{fig5}}
\end{center}
\end{figure}

\begin{figure}
\begin{center}
\includegraphics[angle=0,height=0.9\textheight,width=1.0\textwidth]{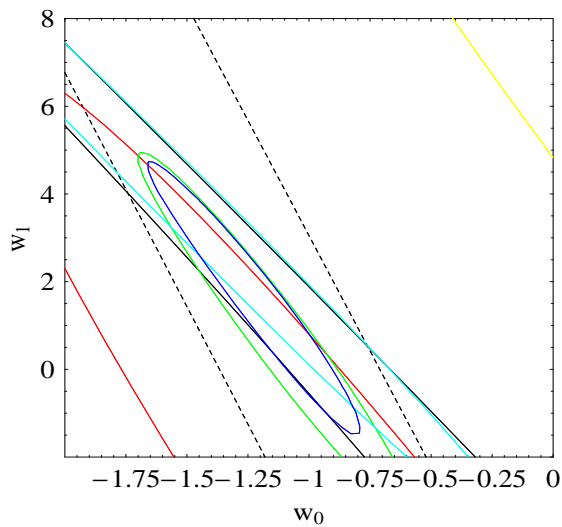}
\caption{The $1\sigma$ joint confidence contours from the
observational datasets in the $w(z)=w_{0}+w_{1}z/(1+z)^2$ model. The
blue contour corresponds to all the datasets. The cyan contour
corresponds to 69 GRBs. The black contour corresponds to the CMB
shift parameter. The green contour corresponds to 182 SNe Ia. The
dashed contour corresponds to BAO. The yellow contour corresponds to
the perturbation growth rate from 2dF Galaxy Redshift Survey.
\label{fig6}}
\end{center}
\end{figure}

\begin{figure}
\begin{center}
\includegraphics[angle=0,height=1.0\textheight,width=1.0\textwidth]{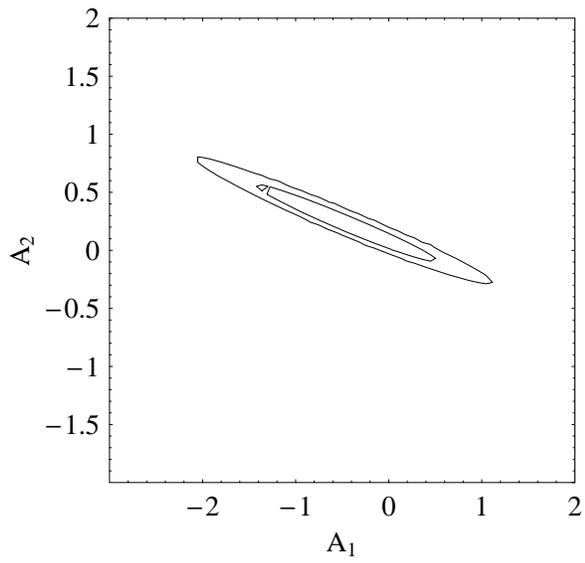}
\caption{The $1\sigma$ and $2\sigma$ joint confidence contours for
$(A_1 ,A_2)$ from all the observational data in the model of
equation (28). \label{fig7}}
 \end{center}
\end{figure}

\begin{figure}
\begin{center}
\includegraphics[angle=0, width=0.5\textwidth]{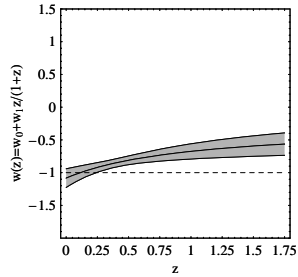}
\includegraphics[angle=0, width=0.5\textwidth]{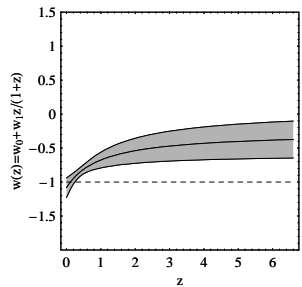}
\caption{The evolution of $w(z)$ by fitting the model
$w(z)=w_{0}+w_{1}z/(1+z)$ to all the observational data. The solid
line represents the reconstructed $w(z)$. The shaded region shows
the $1\sigma$ error. We can constrain the evolution of $w(z)$ up to
$z>6.0$ using GRBs (bottom panel). \label{fig8}}
\end{center}
\end{figure}

\begin{figure}
\begin{center}
\includegraphics[angle=0, width=0.5\textwidth]{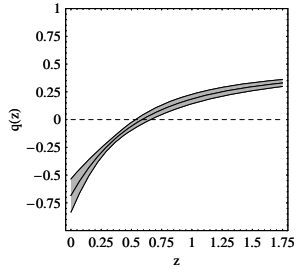}
\includegraphics[angle=0, width=0.5\textwidth]{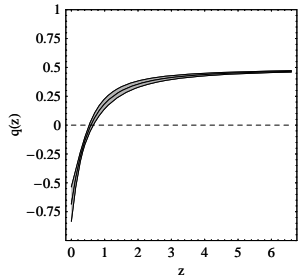}
\caption{The evolution of $q(z)$ by fitting the model
$w(z)=w_{0}+w_{1}z/(1+z)$ to all the observational data. The solid
line is plotted by using the best fitting parameters. The shaded
region shows the $1\sigma$ error. We can reconstruct $q(z)$ up to
$z>6.0$ using GRBs (bottom panel). \label{fig9}}
\end{center}
\end{figure}

\begin{figure}
\begin{center}
\includegraphics[angle=0, width=0.5\textwidth]{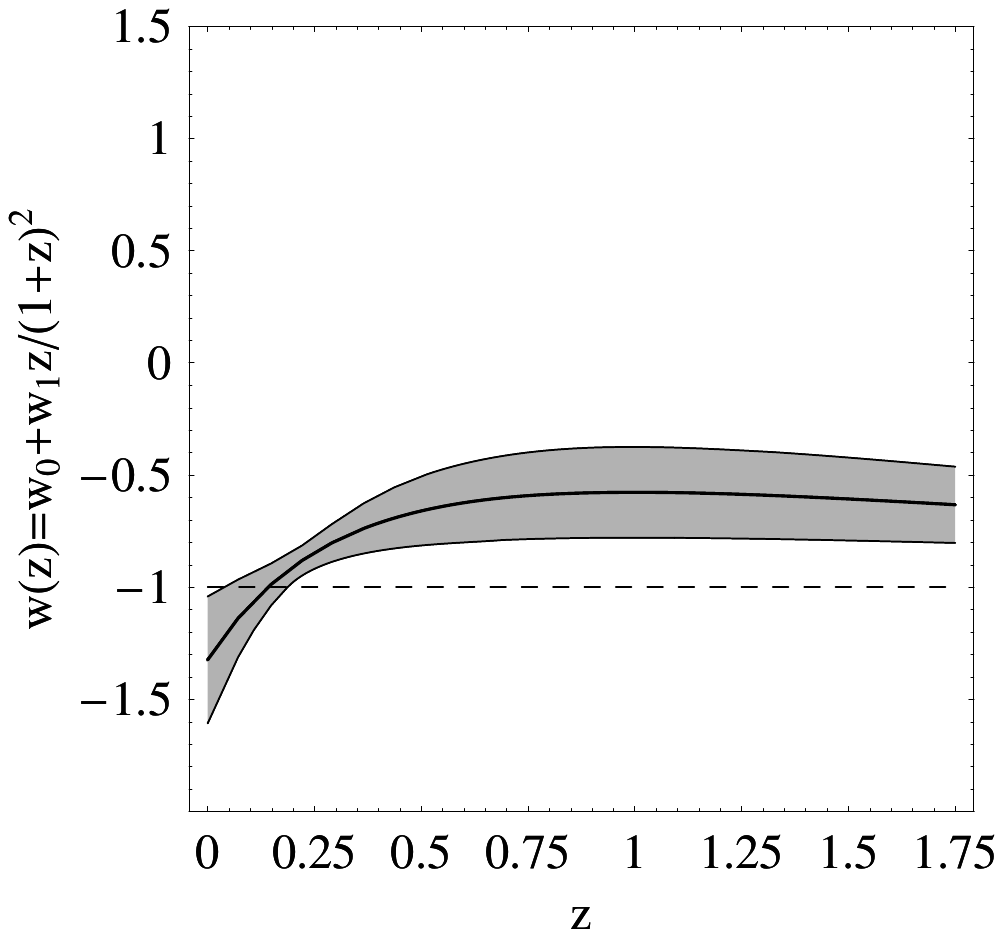}
\includegraphics[angle=0, width=0.5\textwidth]{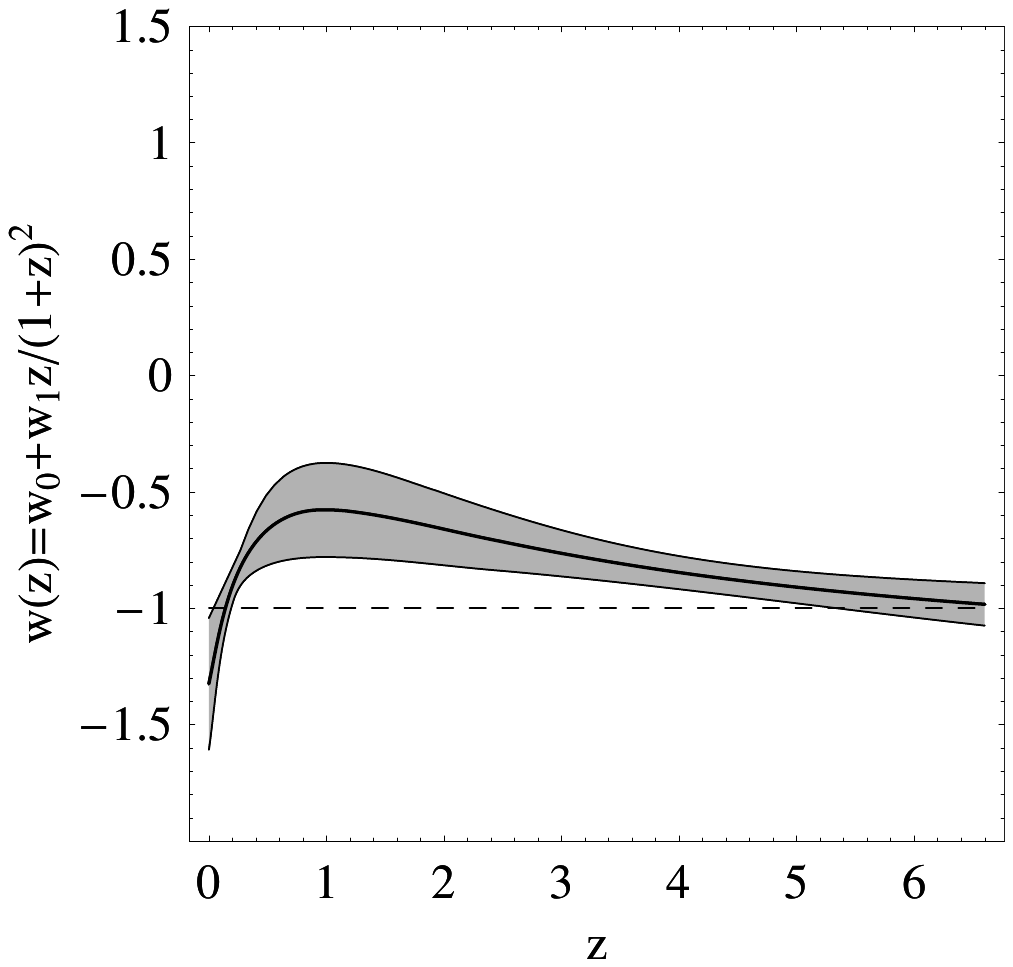}
\caption{Same as Fig.8 but fitting the model
$w(z)=w_{0}+w_{1}z/(1+z)^2$ to all the observational data.
\label{fig10}}
\end{center}
\end{figure}

\begin{figure}
\begin{center}
\includegraphics[angle=0, width=0.5\textwidth]{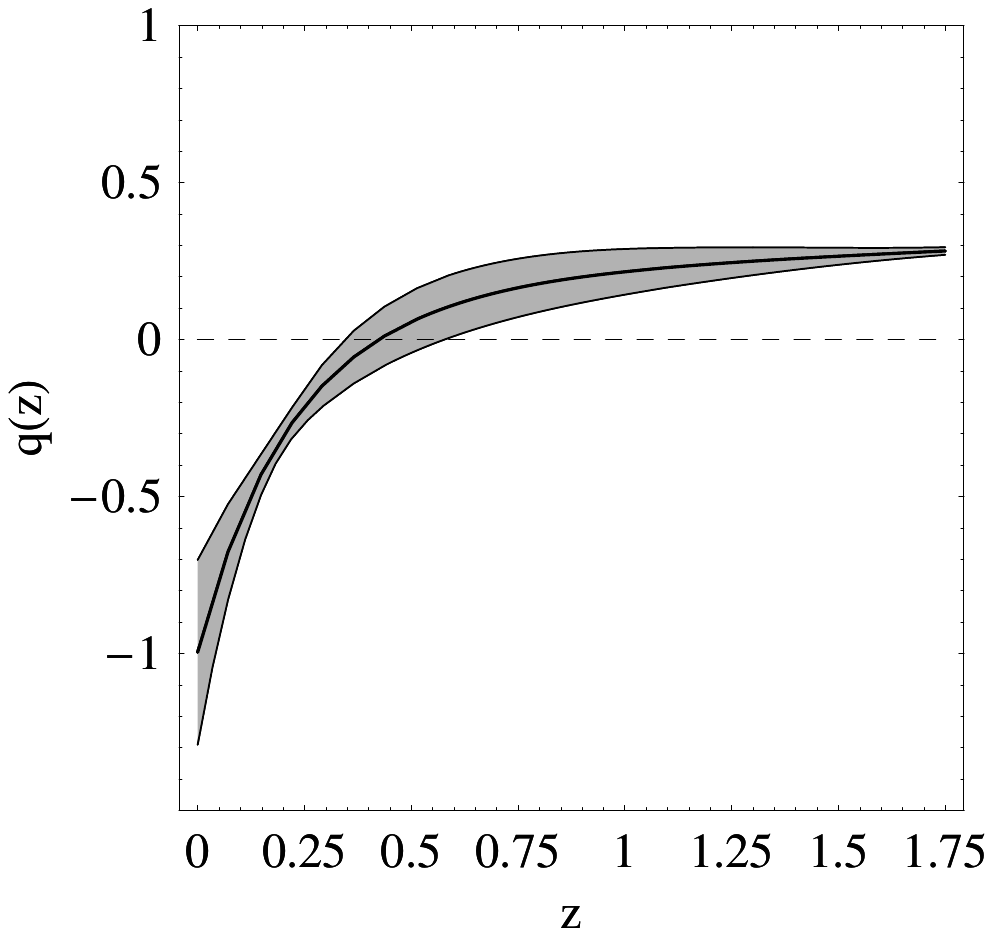}
\includegraphics[angle=0, width=0.5\textwidth]{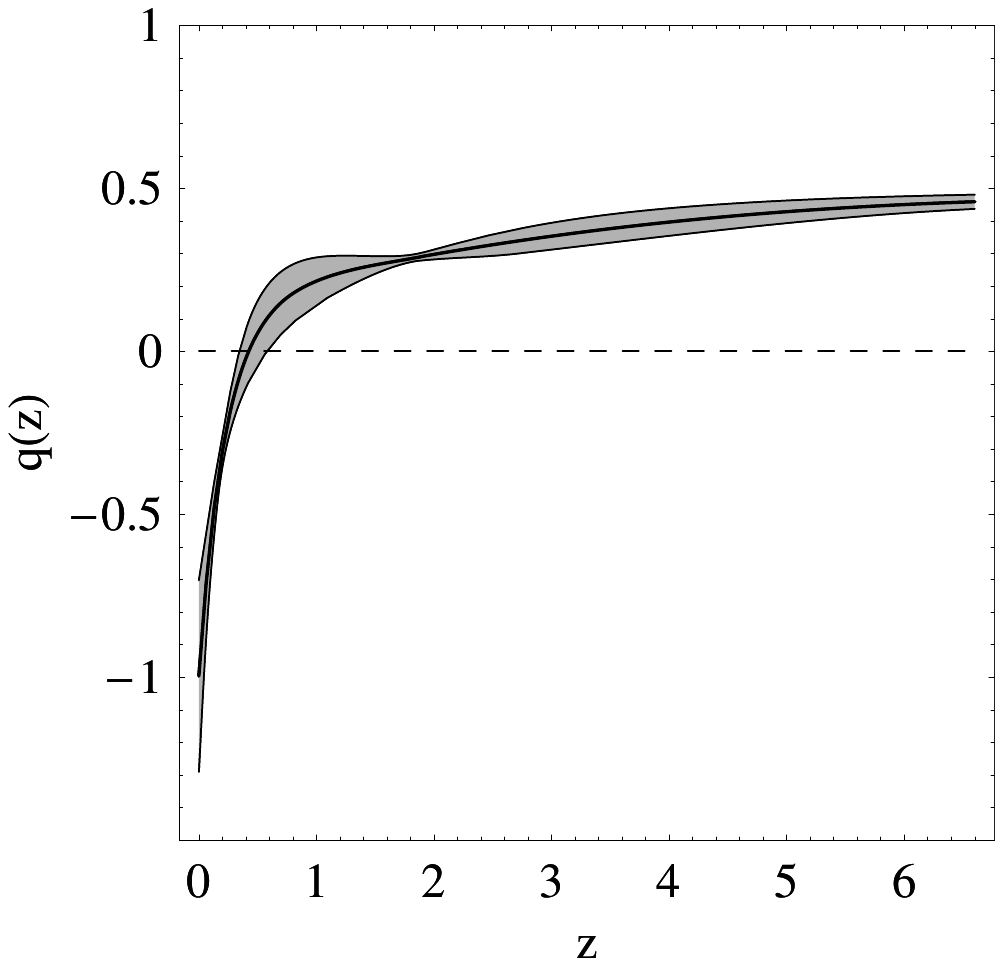}
\caption{Same as Fig.9 but fitting the model
$w(z)=w_{0}+w_{1}z/(1+z)^2$ to all the observational data.
\label{fig11}}
\end{center}
\end{figure}

\begin{figure}
\begin{center}
\includegraphics[angle=0, width=0.5\textwidth]{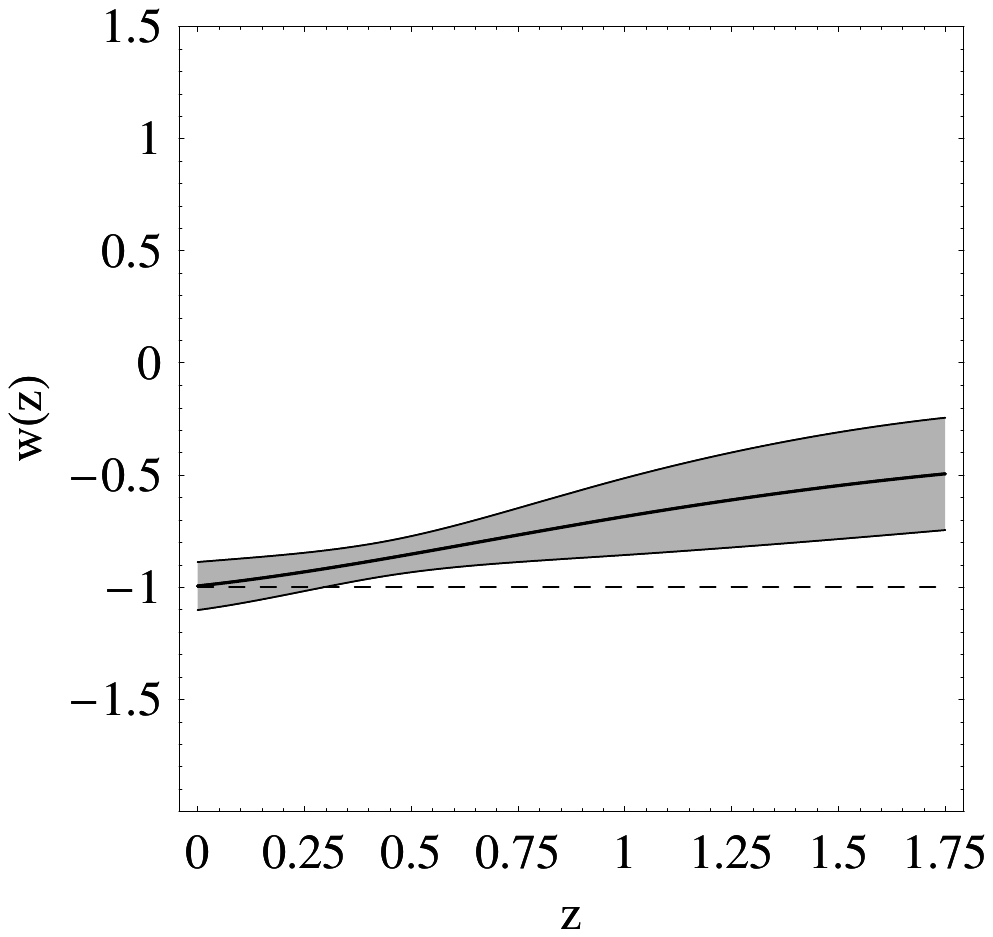}
\includegraphics[angle=0, width=0.5\textwidth]{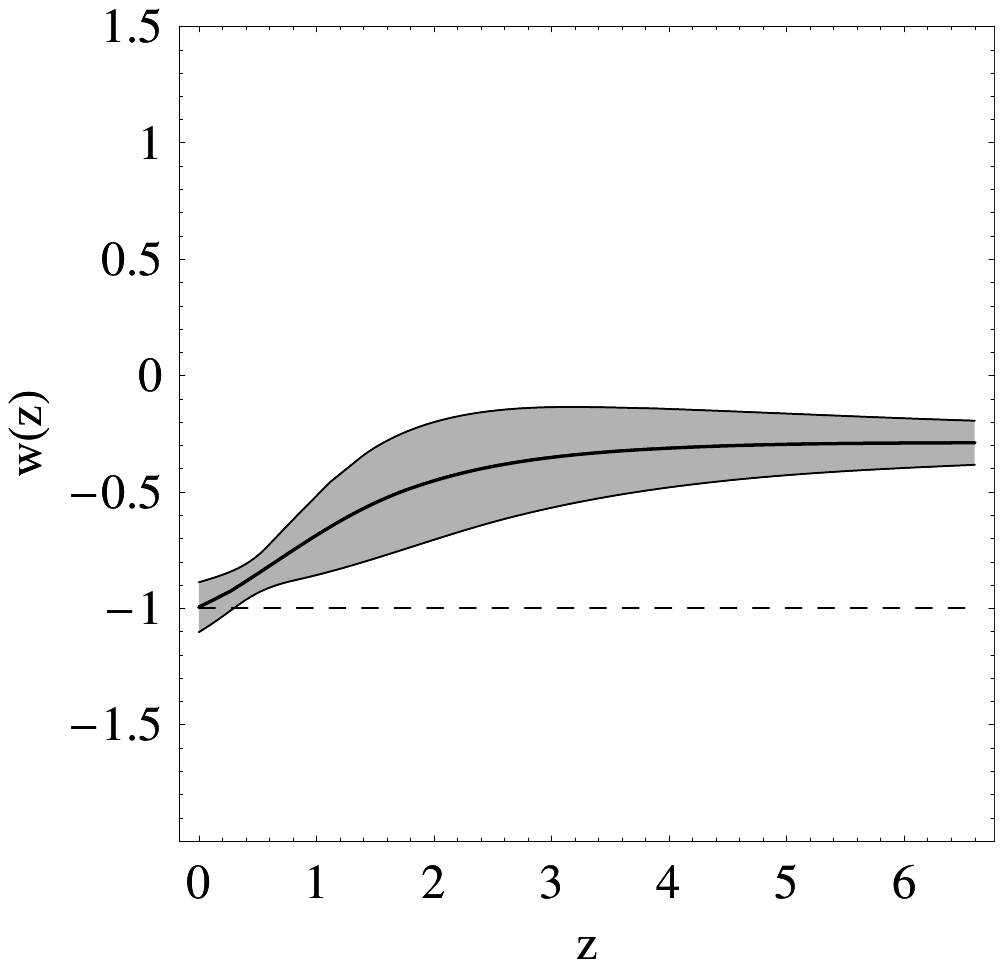}
\caption{Same as Fig.8 but fitting the model of equation (28) to all
the observational data. \label{fig12}}
\end{center}
\end{figure}

\begin{figure}
\begin{center}
\includegraphics[angle=0, width=0.5\textwidth]{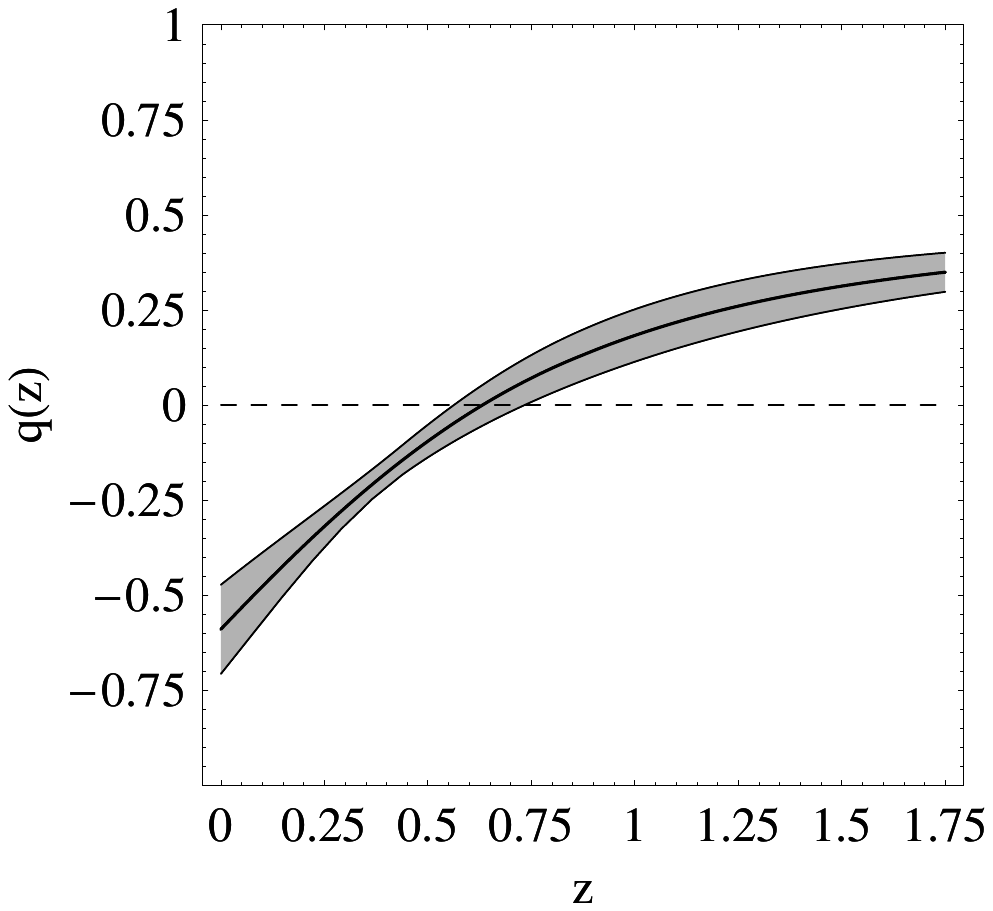}
\includegraphics[angle=0, width=0.5\textwidth]{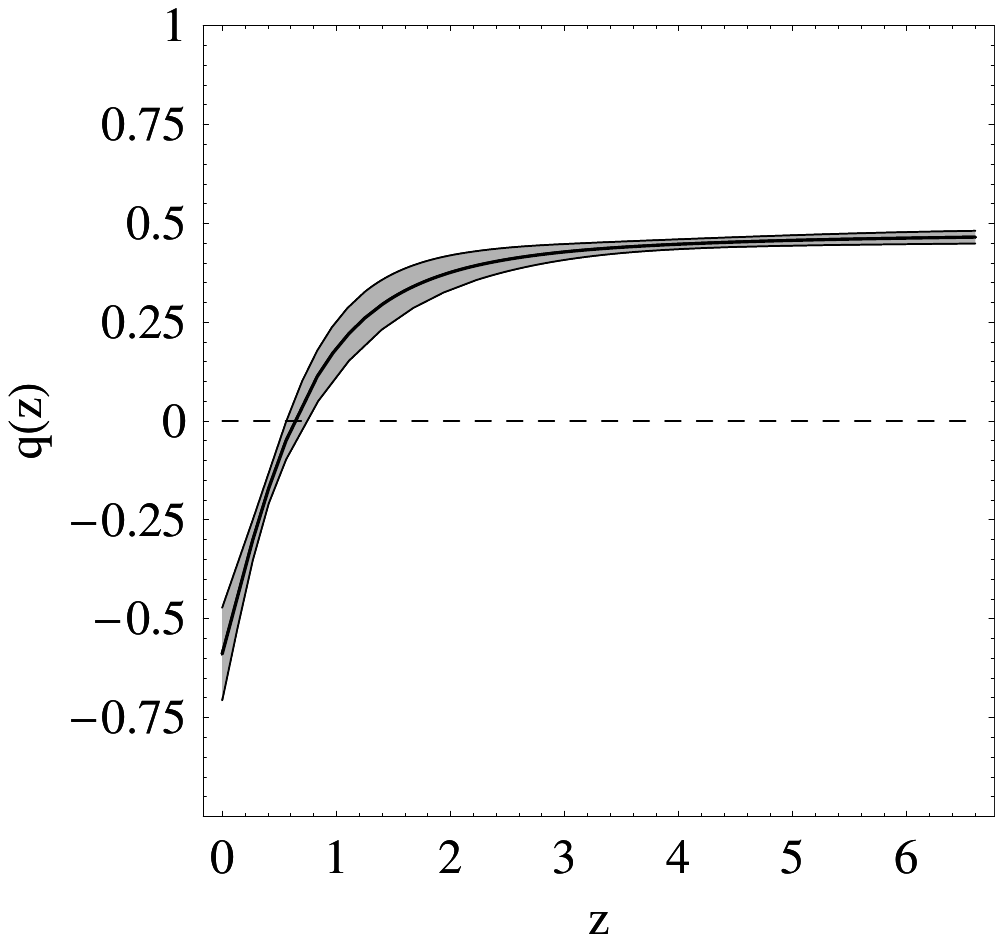}
\caption{Same as Fig.9 but fitting the model of equation (28) to all
the observational data. \label{fig13}}
\end{center}
\end{figure}

\end{document}